# Rate Adaptation via Link-Layer Feedback for Goodput Maximization over a Time-Varying Channel

Rohit Aggarwal, Philip Schniter, and C. Emre Koksal

### Abstract

We consider adapting the transmission rate to maximize the goodput, i.e., the amount of data transmitted without error, over a continuous Markov flat-fading wireless channel. In particular, we consider schemes in which transmitter channel state is inferred from degraded causal error-rate feedback, such as packet-level ACK/NAKs in an automatic repeat request (ARQ) system. In such schemes, the choice of transmission rate affects not only the subsequent goodput but also the subsequent feedback, implying that the optimal rate schedule is given by a partially observable Markov decision process (POMDP). Because solution of the POMDP is computationally impractical, we consider simple suboptimal greedy rate assignment and show that the optimal scheme would itself be greedy if the error-rate feedback was non-degraded. Furthermore, we show that greedy rate assignment using non-degraded feedback yields a total goodput that upper bounds that of optimal rate assignment using degraded feedback. We then detail the implementation of the greedy scheme and propose a reduced-complexity greedy scheme that adapts the transmission rate only once per block of packets. We also investigate the performance of the schemes numerically, and show that the proposed greedy scheme achieves steady-state goodputs that are reasonably close to the upper bound on goodput calculated using non-degraded feedback. A similar improvement is obtained in steady-state goodput, drop rate, and average buffer occupancy in the presence of data buffers. We also investigate an upper bound on the performance of optimal rate assignment for a discrete approximation of the channel and show that such quantization leads to a significant loss in achievable goodput.

*Keywords*: Communication systems, adaptive modulation, ARQ, cross-layer strategies.

## I. INTRODUCTION

Channel variability is common to all wireless communications due to factors such as fading, mobility, and multiuser interference. One way to combat the detrimental effects of channel variability is through rate adaptation [1]–[15]. The idea is that, based on the predicted channel state, the transmitter optimizes the data rate in an effort to maximize the amount of data communicated without error, i.e., the goodput. For example, when the channel quality is below average, the data rate should be decreased to avoid reception errors, while, when the channel quality is above average, the data rate should be increased to prevent the channel from being underutilized.

Rate adaptation would be relatively straightforward if the transmitter could perfectly predict the channel. In practice, however, maintaining accurate transmitter channel state information is a nontrivial task that can consume valuable resources. In *pilot aided* rate adaptation (e.g., [1]–[4]), the transmitter broadcasts physical-layer pilots that the receiver uses to estimate the channel. The receiver's channel estimates (or quantized versions thereof) are then fed back to the transmitter for use in rate adaptation. Because this approach consumes bandwidth on both the forward and reverse links, there is a strong motivation to consider non-pilot-aided approaches. In this paper, we focus on rate adaptation schemes that infer the channel state by monitoring packet acknowledgments/ negative-acknowledgments (ACK/NAKs) [5]–[15],

The authors are with the Department of Electrical and Computer Engineering at The Ohio State University, Columbus, OH 43210. Please direct all correspondence to Prof. Phil Schniter, Dept. ECE, 2015 Neil Ave., Columbus OH 43210, e-mail: schniter@ece.osu.edu, phone 614.247.6488, fax 614.292.7596. Rohit Aggarwal and Emre Koksal can be reached at the same address/phone/fax and e-mailed at aggarwar@ece.osu.edu and koksal@ece.osu.edu.

This work was supported by NSF CAREER grant 237037 and the Office of Naval Research grant N00014-07-1-0209. Portions of this work were presented at the 2008 Conference on Information Science and Systems (CISS), Princeton, NJ, Mar. 2008.

i.e., the feedback information used for automatic repeat request (ARQ). Since ARQ feedback is a standard provision of the link layer, its use by the physical layer comes essentially "for free."

From the viewpoint of rate adaptation, though, ARQ feedback is quite different than pilot-aided feedback. While pilot-aided feedback provides an indicator of the *absolute* channel gain, ARQ provides an indicator of the channel gain *relative* to the chosen data rate; an ACK implies that the channel was good enough to support the rate while a NAK implies otherwise. As a result, with ARQ-feedback based rate adaptation, the chosen data rate affects not only the subsequent goodput but also the quality of the subsequent feedback, which in turn will affect future goodputs through future rate assignments. In fact, with ARQ feedback, optimal rate assignment for communication over Markov channels can be recognized as a dynamic program [14], in particular, a partially observable Markov decision process (POMDP) [16].

In this paper, we consider the general problem of adapting the transmission rate via delayed and degraded error-rate feedback in order to maximize long-term expected goodput. In order to circumvent the sub-optimality of finite-state channel approximations [17], we assume a Markov channel indexed by a *continuous* parameter. Because the optimal solution of the POMDP is too difficult to obtain, we consider the use of *greedy* rate adaptation. First we establish that the optimal rate assignment is itself greedy when the error-rate feedback is *not* degraded. Furthermore, we establish that the greedy non-degraded scheme can be used to upper bound the optimal degraded scheme in terms of long-term goodput. Second, we outline a novel implementation of the greedy rate assignment scheme. For the example case of binary (i.e., ACK/NAK) degraded error-rate feedback, a Rayleigh-fading channel, and uncoded QAM modulation, we show (numerically) that the long-term goodput achieved by our greedy rate assignment scheme is close to the upper bound.

Compared to the previous works [5]–[13], which are ad hoc in nature, we take a more structured approach to cross-layer rate adaptation. Compared to the POMDP-based work [14], our work differs in the following key aspects: 1) we employ a continuous-state Markov channel model, 2) we consider delay in the feedback channel, and 3) we propose simpler greedy heuristics, which we study analytically as well as numerically. Though our adaptation objective—goodput maximization—does not explicitly consider the input buffer state,[1] as does the one in [14], we show (numerically) that finite buffer effects (e.g., packet delay and drop rate) are handled gracefully by our greedy algorithms. In fact, one could argue that, since only successfully communicated packets are removed from the input queue, the maximization of *short-term* goodput—our greedy objective—leads simultaneously to the minimization of buffer occupancy.

The paper is organized as follows. In Section II, we outline our system model, and in Section III, we consider optimal rate adaptation and suboptimal greedy approaches. In Section IV, we detail a novel implementation of the greedy rate-assignment scheme, which we then analyze numerically in Section V for the case of uncoded QAM transmission, ACK/NAK feedback, and a Rayleigh-fading channel. We summarize our findings in Section VI.

## II. System Model

We consider a packetized transmission system in which the transmitter receives delayed and degraded feedback on the success of previous packet transmissions (e.g., binary ACK/NAKs), which it uses to adapt the subsequent transmission parameters. In particular, we assume the use of a transmission scheme parameterized by a data rate of $r_t$ bits per packet, where $t$ denotes the packet index. For simplicity, we assume a fixed transmission power and a fixed packet length of $p$ channel uses.

Figure 1 shows the system model. The time-varying wireless channel is modeled by an SNR process $\{\gamma_t\}$, where the SNR $\gamma_t \geqslant 0$ is assumed to be constant over the packet duration. Notice that $\gamma_t$ is not assumed to be a discrete parameter. Since the transmission power is fixed, $\{\gamma_t\}$ is an exogenous process that does not depend on the transmission parameters. The instantaneous packet error rate $\epsilon(r_t, \gamma_t)$ varies with the rate $r_t$ and SNR $\gamma_t$ according to the particular modulation/demodulation scheme in use. The instantaneous goodput $G(r_t, \gamma_t)$, defined as the number of successfully communicated bits per channel use, is then given by

$$G(r_t, \gamma_t) = \big(1 - \epsilon(r_t, \gamma_t)\big)r_t, \tag{1}$$

where $\epsilon(r_t, \gamma_t)$ denotes the error rate at time $t$. The transmitter uses $\hat{\epsilon}_{t-d}$, a degraded version of the ($d \geqslant 1$ delayed) error rate $\epsilon(r_{t-d}, \gamma_{t-d})$ to choose the time-$t$ rate parameter $r_t$. We will assume that, for each $r_t$, the function $\epsilon(r_t, \gamma_t)$ is monotonically decreasing in $\gamma_t$, so that $\gamma_t$ can be uniquely determined given $r_t$ and the true error rate $\epsilon(r_t, \gamma_t)$.

---

[1] For algorithm design, we assume an infinitely back-logged queue.

*Example 1:* As an illustrative example, we now consider uncoded quadrature amplitude modulation (QAM) using a square constellation of size $m$, and minimum-distance decision making. At the link layer, where symbols are grouped to form packets, a fixed number of extra cyclic redundancy check (CRC) bits are appended to each packet for the purpose of error detection. We will assume the probability of undetected error is negligible and the associated ACK/NAK error feedback is sent back to the transmitter over an error-free reverse channel. We will also assume that the number of CRC bits is small compared to the packet size, allowing us to ignore them in goodput calculations.

Under an AWGN channel, and with $\gamma$ describing the ratio of received symbol power to additive noise power, the symbol error rate for minimum-distance decision making is [18, p. 280]

$$1 - \left(1 - 2\left(1 - \frac{1}{\sqrt{m}}\right)Q\left(\sqrt{\frac{3\gamma}{m-1}}\right)\right)^2, \tag{2}$$

where $Q(\cdot)$ denotes the $Q$-function [18]. If we assume that the constellation size is fixed over the packet duration, then the data rate equals $r_t = p\log_2 m_t$ and the packet error rate equals

$$\epsilon(r_t, \gamma_t) = 1 - \left(1 - 2\left(1 - \frac{1}{\sqrt{2^{r_t/p}}}\right)Q\left(\sqrt{\frac{3\gamma_t}{2^{r_t/p}-1}}\right)\right)^{2p}. \tag{3}$$

Plugging (3) into (1) yields the instantaneous goodput expression, which identifies a particular one-to-one mapping between rate $r_t$ and goodput for a fixed SNR $\gamma_t$. Thus, if the SNR was known perfectly, then the goodput could be maximized by appropriate choice of constellation size. Figure 2 plots instantaneous goodput contours versus SNR $\gamma_t$ and constellation size $m_t$ for the case of $p = 100$ symbols per packet. Figure 2 also plots the (unique) goodput-maximizing constellation size as a function of SNR. Here the finite set of allowed constellation choices (and hence rates) is apparent. Note that, for this uncoded communication scheme, the SNR must be relatively high to facilitate rate adaptation; as long as the SNR remains below 14 dB, the goodput-maximizing constellation size remains at $m = 4$ (i.e., QPSK). Coded transmission, on the other hand, could facilitate rate adaptation at lower SNRs.

With ACK/NAK error feedback, the degraded error-rate $\hat{\epsilon}_t$ is a Bernoulli random variable generated from $\epsilon(r_t, \gamma_t)$ according to the conditional probability mass function

$$p\big(\hat{\epsilon}_t = k\big|\epsilon(r_t, \gamma_t)\big) = \begin{cases} \epsilon(r_t, \gamma_t) & k = 1 \\ 1 - \epsilon(r_t, \gamma_t) & k = 0 \\ 0 & \text{else.} \end{cases} \tag{4}$$

$\square$

While the previous example focuses on a particular modulation/demodulation scheme and a particular error feedback model, we emphasize that the principal results in the sequel are general; no particular modulation/demodulation scheme and error feedback model are assumed.

## III. Optimal Rate Adaptation

In this section we formalize the problem of finite horizon goodput maximization. For convenience we assume that process $\{\gamma_t, \hat{\epsilon}_t, r_t\}$ has been initiated at time $t = -\infty$, though we consider only the finite sequence of packet indices $\{0, \ldots, T\}$ for goodput maximization. Also, we use the abbreviation $\epsilon_t \triangleq \epsilon(r_t, \gamma_t)$.

For every packet index $t \geqslant 0$, we assume that the rate controller has access to the (degraded) error-rate feedback $\hat{\boldsymbol{\epsilon}}_{t-d} \triangleq [\ldots, \hat{\epsilon}_{-2}, \hat{\epsilon}_{-1}, \hat{\epsilon}_0, \ldots, \hat{\epsilon}_{t-d}]$, where $d \geqslant 1$ denotes the causal feedback delay. Formally, we consider $\hat{\boldsymbol{\epsilon}}_{t-d}$ to be *degraded* relative to the true error-rate vector $\boldsymbol{\epsilon}_{t-d}$ if

$$\mathrm{E}\{G(r_t, \gamma_t) \mid \hat{\boldsymbol{\epsilon}}_{t-d}, \boldsymbol{r}_{t-d}\} \neq \mathrm{E}\{G(r_t, \gamma_t) \mid \boldsymbol{\epsilon}_{t-d}, \boldsymbol{r}_{t-d}\} \tag{5}$$

$$= \mathrm{E}\{G(r_t, \gamma_t) \mid \boldsymbol{\gamma}_{t-d}\}, \tag{6}$$

where $\boldsymbol{r}_{t-d} \triangleq [\ldots, r_{-2}, r_{-1}, r_0, \ldots, r_{t-d}]$. Equation (6) follows because each SNR $\gamma_k$ in $\boldsymbol{\gamma}_{t-d} \triangleq [\ldots, \gamma_{-2}, \gamma_{-1}, \gamma_0, \ldots,$ can be uniquely determined from the pair $(\epsilon_k, r_k)$.

At packet index $t$, the *optimal* controller uses the degraded error-rate sequence $\hat{\boldsymbol{\epsilon}}_{t-d}$ (as well as knowledge of the previously chosen rates $\boldsymbol{r}_{t-d}$) to choose the rate $r_t$ from a set $\mathcal{R}$ of admissible rates in order to maximize the total expected goodput for the current and remaining packets:

$$r_t^* \triangleq \arg\max_{r_t \in \mathcal{R}} \, \mathrm{E}\left\{ G(r_t, \gamma_t) + \sum_{k=t+1}^{T} G(r_k^*, \gamma_k) \,\middle|\, \hat{\boldsymbol{\epsilon}}_{t-d}, \boldsymbol{r}_{t-d} \right\} \text{ for } t = 0, \ldots, T. \tag{7}$$

The optimal expected sum goodput for packets $\{t, \ldots, T\}$ can then be written (for $t \geqslant 0$) as

$$G_t^*(\hat{\boldsymbol{\epsilon}}_{t-d}, \boldsymbol{r}_{t-d}) \triangleq \mathrm{E}\left\{ \sum_{k=t}^{T} G(r_k^*, \gamma_k) \,\middle|\, \hat{\boldsymbol{\epsilon}}_{t-d}, \boldsymbol{r}_{t-d} \right\}. \tag{8}$$

For a unit[2] delay system (i.e., $d = 1$), the following Bellman equation [19] specifies the associated finite-horizon dynamic programming problem:

$$\begin{aligned} G_t^*(\hat{\boldsymbol{\epsilon}}_{t-1}, \boldsymbol{r}_{t-1}) = \max_{r_t \in \mathcal{R}} \Big\{ &\mathrm{E}\{G(r_t, \gamma_t) \mid \hat{\boldsymbol{\epsilon}}_{t-1}, \boldsymbol{r}_{t-1}\} \\ &+ \mathrm{E}\{G_{t+1}^*([\hat{\boldsymbol{\epsilon}}_{t-1}, \hat{\epsilon}_t], [\boldsymbol{r}_{t-1}, r_t]) \mid \hat{\boldsymbol{\epsilon}}_{t-1}, \boldsymbol{r}_{t-1}\} \Big\}, \end{aligned} \tag{9}$$

where the second expectation is over $\hat{\epsilon}_t$. The solution to this problem is sometimes referred to as a partially observable Markov decision process (POMDP) [16].

For practical horizons $T$, optimal rate selection based on (9) is computationally impractical, in part due to the continuous-state nature of the channel.[3] In fact, it is known that POMDPs are PSPACE-complete, i.e., they require both complexity and memory that grow exponentially with the horizon $T$ [20]. For an intuitive understanding of this phenomenon, notice from (9) that the solution of the rate assignment problem at every time $t$ depends on the optimal rate assignments up to time $t - 1$. But, because both terms on the right side of (9) are dependent on $r_t$, the solution of the rate assignment problem at time $t$ also depends on the solution of the rate assignment problem at time $t + 1$, which in turn depends on the solution of the rate assignment problem at time $t + 2$, and so on. Consequently, the much simpler *greedy* rate assignment scheme

$$\hat{r}_t \triangleq \arg\max_{r_t \in \mathcal{R}} \, \mathrm{E}\{G(r_t, \gamma_t) \mid \hat{\boldsymbol{\epsilon}}_{t-d}, \boldsymbol{r}_{t-d}\} \text{ for } t = 0, \ldots, T, \tag{10}$$

is suboptimal.

The question of principal interest is then: *What is the loss in goodput with the greedy scheme (10) relative to the optimal scheme (7)?* Since it is too difficult to compute the optimal goodput (which depends on the optimal rate assignment $\boldsymbol{r}_T^*$), we instead compare the greedy scheme (10) to an *upper bound* on the optimal goodput. To establish the upper bound, we show that greedy rate assignment using *non-degraded* error-rate feedback yields a total goodput that is no less than that of optimal rate assignment using degraded error-rate feedback. While the latter is difficult to compute, the former is not.

---

[2] For the $d > 1$ case, the Bellman equation is more complicated, and so we omit it for brevity.

[3] Though a quantized channel approximation could—with few enough states—yield a practical POMDP solution, we show in Section V that channel quantization leads to significant loss in goodput.

We now detail the rate assignment scheme that leads to our total-goodput upper bound. At packet index $t$, consider the rate assignment that maximizes the total expected goodput for the current and remaining packets using knowledge of the non-degraded feedback $\boldsymbol{\epsilon}_{t-d}$:

$$r_t^{\text{cg}} \triangleq \arg \max_{r_t \in \mathcal{R}} \ \mathrm{E} \left\{ G(r_t, \gamma_t) + \sum_{k=t+1}^{T} G(r_k^{\text{cg}}, \gamma_k) \ \middle| \ \boldsymbol{\epsilon}_{t-d}, \boldsymbol{r}_{t-d} \right\} \ \text{ for } \ t = 0, \ldots, T. \tag{11}$$

We refer to this scheme as the *causal genie*. Note that (11) differs from (7) only in that $\boldsymbol{\epsilon}_{t-d}$ is used in place of $\hat{\boldsymbol{\epsilon}}_{t-d}$. Because $\boldsymbol{\gamma}_{t-d}$ can be uniquely determined from $(\boldsymbol{\epsilon}_{t-d}, \boldsymbol{r}_{t-d})$, the causal genie can also be written as

$$r_t^{\text{cg}} = \arg \max_{r_t \in \mathcal{R}} \ \mathrm{E} \left\{ G(r_t, \gamma_t) + \sum_{k=t+1}^{T} G(r_k^{\text{cg}}, \gamma_k) \ \middle| \ \boldsymbol{\gamma}_{t-d} \right\} \ \text{ for } \ t = 0, \ldots, T. \tag{12}$$

Since the choice of $\{r_k^{\text{cg}}\}_{k=t+1}^{T}$ will not depend on the choice of $r_t$, the optimal expected sum goodput for packets $\{t, \ldots, T\}$ can be written (for $t \geqslant 0$) as

$$G_t^{\text{cg}}(\boldsymbol{\epsilon}_{t-d}, \boldsymbol{r}_{t-d}) \triangleq \max_{r_t \in \mathcal{R}} \ \mathrm{E} \left\{ G(r_t, \gamma_t) + \sum_{k=t+1}^{T} G(r_k^{\text{cg}}, \gamma_k) \ \middle| \ \boldsymbol{\gamma}_{t-d} \right\} \tag{13}$$

$$= \mathrm{E} \left\{ \sum_{k=t+1}^{T} G(r_k^{\text{cg}}, \gamma_k) \ \middle| \ \boldsymbol{\gamma}_{t-d} \right\} + \max_{r_t \in \mathcal{R}} \mathrm{E}\{G(r_t, \gamma_t) \mid \boldsymbol{\gamma}_{t-d}\}, \tag{14}$$

which shows that *optimal rate assignment under non-degraded causal error-rate feedback can be accomplished greedily*. In other words,

$$r_t^{\text{cg}} = \arg \max_{r_t \in \mathcal{R}} \ \mathrm{E}\{G(r_t, \gamma_t) \mid \boldsymbol{\gamma}_{t-d}\} \tag{15}$$

$$= \arg \max_{r_t \in \mathcal{R}} \ \mathrm{E}\{G(r_t, \gamma_t) \mid \boldsymbol{\epsilon}_{t-d}, \boldsymbol{r}_{t-d}\}. \tag{16}$$

We now establish that the causal genie controller upper bounds the optimal controller with degraded error-rate feedback in the sense of total goodput. Though the result may be intuitive, the proof provides insight into the relationship between degradation of the feedback and reduction of the total expected goodput.

*Lemma 1:* Given arbitrary past rates $\boldsymbol{r}_{-d}$ and corresponding degraded error-rate feedback $\hat{\boldsymbol{\epsilon}}_{-d}$, the expected total goodput for optimal rate allocation under degraded feedback is no higher than the expected total goodput for the causal-genie rate allocation under non-degraded feedback, i.e.,

$$\mathrm{E} \left\{ \sum_{t=0}^{T} G(r_t^*, \gamma_t) \ \middle| \ \hat{\boldsymbol{\epsilon}}_{-d}, \boldsymbol{r}_{-d} \right\} \leqslant \mathrm{E} \left\{ \sum_{t=0}^{T} G(r_t^{\text{cg}}, \gamma_t) \ \middle| \ \hat{\boldsymbol{\epsilon}}_{-d}, \boldsymbol{r}_{-d} \right\}. \tag{17}$$

*Proof:* For any $t \in \{0, \ldots, T\}$ and any realization of $(\hat{\boldsymbol{\epsilon}}_{t-d}, \boldsymbol{r}_{t-d})$, we can write

$$\mathrm{E}\{G(r_t^*, \gamma_t) \mid \hat{\boldsymbol{\epsilon}}_{t-d}, \boldsymbol{r}_{t-d}\} \leqslant \max_{r_t \in \mathcal{R}} \ \mathrm{E}\{G(r_t, \gamma_t) \mid \hat{\boldsymbol{\epsilon}}_{t-d}, \boldsymbol{r}_{t-d}\} \tag{18}$$

$$= \max_{r_t \in \mathcal{R}} \ \mathrm{E} \left\{ \mathrm{E}\{G(r_t, \gamma_t) \mid \hat{\boldsymbol{\epsilon}}_{t-d}, \boldsymbol{r}_{t-d}, \boldsymbol{\epsilon}_{t-d}\} \ \middle| \ \hat{\boldsymbol{\epsilon}}_{t-d}, \boldsymbol{r}_{t-d} \right\} \tag{19}$$

$$\leqslant \mathrm{E} \left\{ \max_{r_t \in \mathcal{R}} \ \mathrm{E}\{G(r_t, \gamma_t) \mid \hat{\boldsymbol{\epsilon}}_{t-d}, \boldsymbol{r}_{t-d}, \boldsymbol{\epsilon}_{t-d}\} \ \middle| \ \hat{\boldsymbol{\epsilon}}_{t-d}, \boldsymbol{r}_{t-d} \right\} \tag{20}$$

$$= \mathrm{E} \left\{ \max_{r_t \in \mathcal{R}} \ \mathrm{E}\{G(r_t, \gamma_t) \mid \boldsymbol{\gamma}_{t-d}\} \ \middle| \ \hat{\boldsymbol{\epsilon}}_{t-d}, \boldsymbol{r}_{t-d} \right\} \tag{21}$$

$$= \mathrm{E}\{G(r_t^{\text{cg}}, \gamma_t) \mid \hat{\boldsymbol{\epsilon}}_{t-d}, \boldsymbol{r}_{t-d}\}, \tag{22}$$

where (18) follows since $r_t^*$ is chosen to maximize the long term goodput—not the instantaneous goodput; (20) follows since $\max_{r_t} \mathrm{E}\{f(r_t)\} \leqslant \mathrm{E}\{\max_{r_t} f(r_t)\}$ for any $f(\cdot)$; (21) follows by definition of degraded feedback; and (22) follows by definition of the greedy genie. Taking the expectation over $(\hat{\boldsymbol{\epsilon}}_{-d}, \boldsymbol{r}_{-d})$, conditional on $(\hat{\boldsymbol{\epsilon}}_{-d}, \boldsymbol{r}_{-d})$, we find

$$\mathrm{E}\{G(r_t^*, \gamma_t) \mid \hat{\boldsymbol{\epsilon}}_{-d}, \boldsymbol{r}_{-d}\} \leqslant \mathrm{E}\{G(r_t^{cg}, \gamma_t) \mid \hat{\boldsymbol{\epsilon}}_{-d}, \boldsymbol{r}_{-d}\}. \tag{23}$$

Finally, summing both sides of (23) over $t = \{0, \dots, T\}$ yields (17). ∎

In Section IV we study the greedy rate assignment scheme (10) in depth. Then, in Section V, we study (numerically) the particular case in which $\{\hat{\epsilon}_t\}_{t \geqslant 0}$ is constructed from link-layer ACK/NAKs.

## IV. The Greedy Rate Adaptation Algorithm

In this section, we detail the implementation of greedy rate assignment (10) assuming continuous Markov SNR variation and conditionally independent error-rate estimates. In Section IV-A, we detail a procedure for packet-rate adaptation, while in Section IV-B, we consider adapting the rate once per block of $n$ packets.

### A. Packet-Rate Algorithm

Assuming a feedback delay of $d \geqslant 1$ packets, the greedy rate assignment (10) can be rewritten as

$$\hat{r}_t = \arg\max_{r_t \in \mathcal{R}} \int G(r_t, \gamma_t) p(\gamma_t \mid \hat{\boldsymbol{\epsilon}}_{-d}, \boldsymbol{r}_{t-d}) \ d\gamma_t \ \text{ for } t = 0, \dots, T. \tag{24}$$

We now derive a recursive implementation of the greedy rate assignment (24).

Expanding the inferred SNR distribution via Bayes rule, we find

$$p(\gamma_t \mid \hat{\boldsymbol{\epsilon}}_{t-d}, \boldsymbol{r}_{t-d})$$
$$= \int p(\gamma_t \mid \gamma_{t-d}, \hat{\boldsymbol{\epsilon}}_{t-d}, \boldsymbol{r}_{t-d}) p(\gamma_{t-d} \mid \hat{\boldsymbol{\epsilon}}_{t-d}, \boldsymbol{r}_{t-d}) \ d\gamma_{t-d} \tag{25}$$
$$= \int p(\gamma_t \mid \gamma_{t-d}) p(\gamma_{t-d} \mid \hat{\boldsymbol{\epsilon}}_{t-d}, \boldsymbol{r}_{t-d}) \ d\gamma_{t-d}, \tag{26}$$

where we used the assumption of Markov SNR variation to write (26). Furthermore,

$$p(\gamma_{t-d} \mid \hat{\boldsymbol{\epsilon}}_{t-d}, \boldsymbol{r}_{t-d})$$
$$= p(\gamma_{t-d} \mid \hat{\epsilon}_{t-d}, \hat{\boldsymbol{\epsilon}}_{t-d-1}, \boldsymbol{r}_{t-d}) \tag{27}$$
$$= \frac{p(\hat{\epsilon}_{t-d} \mid \gamma_{t-d}, \hat{\boldsymbol{\epsilon}}_{t-d-1}, \boldsymbol{r}_{t-d}) p(\gamma_{t-d} \mid \hat{\boldsymbol{\epsilon}}_{t-d-1}, \boldsymbol{r}_{t-d})}{\int p(\hat{\epsilon}_{t-d} \mid \gamma'_{t-d}, \hat{\boldsymbol{\epsilon}}_{t-d-1}, \boldsymbol{r}_{t-d}) p(\gamma'_{t-d} \mid \hat{\boldsymbol{\epsilon}}_{t-d-1}, \boldsymbol{r}_{t-d}) \ d\gamma'_{t-d}} \tag{28}$$
$$= \frac{p(\hat{\epsilon}_{t-d} \mid \epsilon(r_{t-d}, \gamma_{t-d}), \hat{\boldsymbol{\epsilon}}_{t-d-1}) p(\gamma_{t-d} \mid \hat{\boldsymbol{\epsilon}}_{t-d-1}, \boldsymbol{r}_{t-d-1})}{\int p(\hat{\epsilon}_{t-d} \mid \epsilon(r_{t-d}, \gamma'_{t-d}), \hat{\boldsymbol{\epsilon}}_{t-d-1}) p(\gamma'_{t-d} \mid \hat{\boldsymbol{\epsilon}}_{t-d-1}, \boldsymbol{r}_{t-d-1}) \ d\gamma'_{t-d}}. \tag{29}$$

With conditionally independent error estimates (i.e., $p(\hat{\epsilon}_t \mid \epsilon_t, \hat{\epsilon}_{t-1}) = p(\hat{\epsilon}_t \mid \epsilon_t)$), this becomes

$$p(\gamma_{t-d} \mid \hat{\boldsymbol{\epsilon}}_{t-d}, \boldsymbol{r}_{t-d})$$
$$= \frac{p(\hat{\epsilon}_{t-d} \mid \epsilon(r_{t-d}, \gamma_{t-d})) p(\gamma_{t-d} \mid \hat{\boldsymbol{\epsilon}}_{t-d-1}, \boldsymbol{r}_{t-d-1})}{\int p(\hat{\epsilon}_{t-d} \mid \epsilon(r_{t-d}, \gamma'_{t-d})) p(\gamma'_{t-d} \mid \hat{\boldsymbol{\epsilon}}_{t-d-1}, \boldsymbol{r}_{t-d-1}) \ d\gamma'_{t-d}}. \tag{30}$$

Similar to (26), we can also write

$$p(\gamma_{t-d+1} \mid \hat{\boldsymbol{\epsilon}}_{t-d}, \boldsymbol{r}_{t-d})$$
$$= \int p(\gamma_{t-d+1} \mid \gamma_{t-d}, \hat{\boldsymbol{\epsilon}}_{t-d}, \boldsymbol{r}_{t-d})$$
$$\times p(\gamma_{t-d} \mid \hat{\boldsymbol{\epsilon}}_{t-d}, \boldsymbol{r}_{t-d}) \ d\gamma_{t-d} \tag{31}$$
$$= \int p(\gamma_{t-d+1} \mid \gamma_{t-d}) p(\gamma_{t-d} \mid \hat{\boldsymbol{\epsilon}}_{t-d}, \boldsymbol{r}_{t-d}) \ d\gamma_{t-d}. \tag{32}$$

Equations (26), (30), and (32) lead to the following recursive implementation of the greedy rate assignment (24). Assuming the availability[4] of $p(\gamma_{t-d} \mid \hat{\boldsymbol{\epsilon}}_{t-d-1}, \boldsymbol{r}_{t-d-1})$ when calculating $r_t$, the rate assignment procedure for packet indices $t = 0, \ldots, T$ is:

1) easure $\hat{\epsilon}_{t-d}$, compute $p(\hat{\epsilon}_{t-d} \mid \epsilon(r_{t-d}, \gamma_{t-d}))$ as a function of $\gamma_{t-d}$, and then calculate the distribution $p(\gamma_{t-d} \mid \hat{\epsilon}_{t-d}, \boldsymbol{r}_{t-d})$ using (30).
2) Calculate $p(\gamma_t \mid \hat{\boldsymbol{\epsilon}}_{t-d}, \boldsymbol{r}_{t-d})$ using the Markov prediction step (26).
3) Calculate $\hat{r}_t$ via (24).
4) If[5] $d > 1$, then calculate $p(\gamma_{t-d+1} \mid \hat{\boldsymbol{\epsilon}}_{t-d}, \boldsymbol{r}_{t-d})$ via (32) for use in the next iteration.

### B. Block-Rate Algorithm

Since it may be impractical for the transmitter to adapt the rate on a per-packet basis, we now propose a modification of the algorithm detailed in Section IV-A that adapts the rate only once per block of $n$ packets. The main idea behind our block-rate algorithm is that the SNR $\{\gamma_t\}$ and error-rate estimates $\{\hat{\epsilon}_t\}$ are treated as if they were *constant* over the block, thereby allowing a straightforward application of the method from Section IV-A. Though this treatment is suboptimal, our intention is to trade performance for reduced complexity.

The details of our block-rate algorithm are now given. Denoting the block index by $i$, the block versions of the degraded error-rate estimate and SNR are defined as

$$\underline{\hat{\epsilon}}_i \triangleq \frac{1}{n} \sum_{t=in}^{(i+1)n-1} \hat{\epsilon}_t \tag{33}$$

$$\underline{\gamma}_i \triangleq \gamma_{in+\lfloor n/2 \rfloor}, \tag{34}$$

and the assigned rates $\{r_t\}$ are related to the calculated rates $\{\underline{r}_i\}$ as

$$r_t = \underline{r}_{\lfloor t/n \rfloor}. \tag{35}$$

Notice that, when $n = 1$, the block-rate quantities reduce to the packet-rate quantities, i.e., $\underline{\hat{\epsilon}}_i = \hat{\epsilon}_i$, $\underline{\gamma}_i = \gamma_i$, and $\underline{r}_i = r_i$.

Borrowing the packet-rate adaptation approach from Section IV-A, the block-rate greedy implementation goes as follows. Here, we use $d$ to denote the delay in *blocks*. Assuming the availability[6] of $p(\underline{\gamma}_{i-d} \mid \hat{\underline{\epsilon}}_{i-d-1}, \underline{r}_{i-d-1})$ when calculating $\underline{r}_i$, the rate assignment procedure for block indices $i = 0, \ldots, \lceil T/n \rceil$ is:

1) easure $\{\hat{\epsilon}_t\}_{t=(i-d)n}^{(i-d+1)n-1}$, compute $\hat{\underline{\epsilon}}_{i-d}$ via (33), compute $p(\hat{\underline{\epsilon}}_{i-d} \mid \epsilon(\underline{r}_{i-d}, \underline{\gamma}_{i-d}))$ as a function of $\underline{\gamma}_{i-d}$, and then calculate the inferred SNR distribution $p(\underline{\gamma}_{i-d} \mid \hat{\underline{\epsilon}}_{i-d}, \underline{r}_{i-d})$ using

$$p(\underline{\gamma}_{i-d} \mid \hat{\underline{\epsilon}}_{i-d}, \underline{r}_{i-d})$$
$$= \frac{p(\hat{\underline{\epsilon}}_{i-d} \mid \epsilon(\underline{r}_{i-d}, \underline{\gamma}_{i-d})) p(\underline{\gamma}_{i-d} \mid \hat{\underline{\epsilon}}_{i-d-1}, \underline{r}_{i-d-1})}{\int p(\hat{\underline{\epsilon}}_{i-d} \mid \epsilon(\underline{r}_{i-d}, \underline{\gamma}'_{i-d})) p(\underline{\gamma}'_{i-d} \mid \hat{\underline{\epsilon}}_{i-d-1}, \underline{r}_{i-d-1}) \ d\underline{\gamma}'_{i-d}}. \tag{36}$$

2) Calculate $p(\underline{\gamma}_i \mid \hat{\underline{\epsilon}}_{i-d}, \underline{r}_{i-d})$ using the Markov prediction step

$$p(\underline{\gamma}_i \mid \hat{\underline{\epsilon}}_{i-d}, \underline{r}_{i-d})$$
$$= \int p(\underline{\gamma}_i \mid \underline{\gamma}_{i-d}) p(\underline{\gamma}_{i-d} \mid \hat{\underline{\epsilon}}_{i-d}, \underline{r}_{i-d}) \ d\underline{\gamma}_{i-d}. \tag{37}$$

---

[4] For the initial packet indices $t \in \{0, \ldots, d\}$, if the pdf $p(\gamma_{t-d} \mid \hat{\epsilon}_{t-d-1}, \boldsymbol{r}_{t-d-1})$ is unknown, then we suggest to use the prior $p(\gamma_{t-d})$ in its place.

[5] Notice that, if $d = 1$, then $p(\gamma_{t-d+1} \mid \hat{\epsilon}_{t-d}, \boldsymbol{r}_{t-d})$ was already computed in step 2).

[6] For the initial block indices $i \in \{0, \ldots, d\}$, if the pdf $p(\underline{\gamma}_{i-d} \mid \hat{\underline{\epsilon}}_{i-d-1}, \underline{r}_{i-d-1})$ is unknown, then we suggest to use the prior $p(\underline{\gamma}_{i-d})$ in its place.

3) Calculate $\underline{\hat{r}}_i$ via

$$\underline{\hat{r}}_i = \arg\max_{\underline{r}_i \in \mathcal{R}} \int G(\underline{r}_i, \underline{\gamma}_i) p(\underline{\gamma}_i \mid \underline{\hat{\epsilon}}_{i-d}, \underline{r}_{i-d}) \, d\underline{\gamma}_i. \tag{38}$$

4) If[7] $d > 1$, then calculate $p(\underline{\gamma}_{i-d+1} \mid \underline{\hat{\epsilon}}_{i-d}, \underline{r}_{i-d})$ as follows for use in the next iteration.

$$\begin{aligned} &p(\underline{\gamma}_{i-d+1} \mid \underline{\hat{\epsilon}}_{i-d}, \underline{r}_{i-d}) \\ &= \int p(\underline{\gamma}_{i-d+1} \mid \underline{\gamma}_{i-d}) p(\underline{\gamma}_{i-d} \mid \underline{\hat{\epsilon}}_{i-d}, \underline{r}_{i-d}) \, d\underline{\gamma}_{i-d}. \end{aligned} \tag{39}$$

As the adaptation-block size $n$ increases, we expect the packet error rate estimate $\underline{\hat{\epsilon}}_i$ to become more accurate (since it is estimated from, e.g., $n$ ACK/NAKs), the SNR model to get less accurate (since a block-fading approximation is being applied to a process that is continuously fading), and the per-packet implementation complexity of the algorithm to decrease.

We note that the block-rate modification proposed here is suboptimal in the sense that the SNR of each packet in a block could have been predicted individually, rather than predicting only the SNR of the packet in the middle of the block. Likewise, individual rates could have been assigned for each packet in the block, rather than a uniform rate for all packets in the block. However, joint optimization of intra-block rates appears to be prohibitively complex and thus goes against our primary motivation for the block-rate algorithm, i.e., simplicity.

Finally, we note that a similar block-rate modification can also be applied to the causal genie scheme (16), which has been recognized as a non-degraded-feedback version of the greedy scheme (10). However, doing so would spoil the total-goodput optimality of the packet-rate causal genie that was identified in Lemma 1.

## V. Numerical Results

We now describe the results of numerical experiments in which we assume uncoded square-QAM modulation, a Gauss-Markov fading channel, and minimum variance unbiased (MVU) estimation of the error-rate, as detailed below. While other examples of modulation, error-rate estimation, and fading could have been employed, we feel that our choices are sufficient to illustrate the essential behaviors of the generic rate adaptation schemes discussed in Sections III and IV.

### A. Setup

For our numerical experiments, we used the uncoded QAM modulation/demodulation scheme described in Example 1, which yields the packet error-rate given in (3). We used squared-integer constellation sizes, i.e., 4-QAM, 9-QAM, 16-QAM, etc. In addition, we used causal degraded error-rate feedback in the form of one ACK/NAK per transmitted packet. Thus, in a block[8] of $n$ packets, there were $n$ ACK/NAKs.

Given this setup, it can be shown that the MVU estimate [21] of the average packet error rate over the $i$-th block can be computed by a simple arithmetic average of the $n$ ACK/NAKs, using $0$ for an ACK and $1$ for a NAK. Notice that this MVU estimate corresponds exactly to the block error-rate estimate $\underline{\hat{\epsilon}}_i$ specified in (33). Furthermore, if $\underline{\epsilon}_i$ denotes the value of the true packet error rate over the $i$-th block, then the number of NAKs per block is Binomial$(n, \underline{\epsilon}_i)$ and the error estimate $\underline{\hat{\epsilon}}_i$ obeys

$$p(\underline{\hat{\epsilon}}_i = \tfrac{k}{n} \mid \underline{\epsilon}_i) = \begin{cases} \binom{n}{k} \underline{\epsilon}_i^k (1 - \underline{\epsilon}_i)^{n-k} & \text{for } k = 0, \ldots, n \\ 0 & \text{else.} \end{cases} \tag{40}$$

Thus, we can calculate $\underline{\epsilon}_i = \epsilon(\underline{r}_i, \underline{\gamma}_i)$ as a function of $\underline{\gamma}_i$ using (3) and plug the results into $p(\underline{\hat{\epsilon}}_i \mid \underline{\epsilon}_i)$ from (40) in order to compute (36).

---

[7] Notice that, if $d = 1$, then $p(\underline{\gamma}_{i-d+1} \mid \underline{\hat{\epsilon}}_{i-d}, \underline{r}_{i-d})$ was already computed in step 2).

[8] The results here also hold for packet-rate adaptation through the choice $n = 1$.

To generate the Markov block-rate SNR process $\{\underline{\gamma}_i\}$, we first generate a packet-rate complex-valued Gauss-Markov "channel gain" [22] process $\{g_t\}$ using

$$g_t = (1-\alpha)g_{t-1} + \alpha w_t, \tag{41}$$

where $\{w_t\}$ is a zero-mean unit-variance white circular Gaussian driving process and $0 \leq \alpha \leq 1$. Notice that $\alpha = 1$ corresponds to i.i.d. gains, whereas $\alpha = 0$ corresponds to a time-invariant gain. We then generate a packet-rate SNR process $\{\gamma_t\}$ by scaling the squared magnitude of $g_t$:

$$\gamma_t = K|g_t|^2. \tag{42}$$

The scaling parameter $K$ in (42) is essential because $\alpha$ affects both the (steady-state) coherence time and the mean-squared value of the gain $\{g_t\}$. Thus, by using the two parameters $K$ and $\alpha$, it is possible to independently control the (steady-state) mean and coherence time of the SNR process $\{\gamma_t\}$. In fact, it can be shown that, for steady-state indices $t$, the SNR $\gamma_t$ is exponentially distributed with mean value $\frac{2K\alpha}{2-\alpha}$.

To evaluate $p(\underline{\gamma}_i \mid \underline{\gamma}_{i-d})$, we first notice from (34) that $p(\underline{\gamma}_i \mid \underline{\gamma}_{i-d}) = p(\gamma_t \mid \gamma_{t-nd})$. Then, from (41), we find that

$$g_t = (1-\alpha)^{nd} g_{t-nd} + \alpha \sum_{j=0}^{nd-1} (1-\alpha)^j w_{t-j}, \tag{43}$$

where $\sum_{j=0}^{nd-1}(1-\alpha)^j w_{t-j} \sim \mathcal{CN}\left(0, \frac{1}{1-(1-\alpha)^2}(1-(1-\alpha)^{2nd})\right)$. From this fact, we show in the Appendix that

$$
\begin{aligned}
p(\gamma_t \mid \gamma_{t-nd}) = {} & \frac{2-\alpha}{2K\alpha(1-(1-\alpha)^{2nd})} \\
& \times \exp\left(\frac{-(\gamma_t + (1-\alpha)^{2nd}\gamma_{t-nd})(2-\alpha)}{2K\alpha(1-(1-\alpha)^{2nd})}\right) \\
& \times I_o\left(\frac{(1-\alpha)^{nd}\sqrt{\gamma_t \gamma_{t-nd}}(2-\alpha)}{K\alpha(1-(1-\alpha)^{2nd})}\right).
\end{aligned}
\tag{44}
$$

### B. Results

Numerical experiments were conducted to investigate the steady-state performance of the greedy algorithm from Section IV relative to three reference schemes: *fixed rate*, *causal genie*, and *noncausal genie*, both with and without finite-buffer constraints at the transmitter. The so-called *fixed-rate* reference scheme chooses the fixed rate (i.e., constellation size) that maximizes expected goodput under the prior SNR distribution, i.e., $\arg\max_{r_t} \int G(r_t, \gamma_t) p(\gamma_t) d\gamma_t$. In the absence of feedback, this fixed rate would be optimal, i.e., total-goodput maximizing. The *causal genie* reference scheme defined in Section III adapts the rate to maximize expected goodput under perfect causal feedback of the error rate $\epsilon_t$ or, equivalently, the SNR $\gamma_t$. As shown in Section III, the goodput attained by the causal genie upper bounds that of optimal rate selection under degraded feedback. However, as the feedback delay $d$ and/or the block size $n$ increases, the causal genie's ability to predict the SNR decreases, and thus its goodput suffers. The so-called *non-causal genie* reference scheme assumes perfect knowledge of SNR $\gamma_t$ for all past, current, and future packets, and uses this information to choose the goodput-maximizing rate. Since this scheme has access to more information than the causal-genie and greedy algorithms, it upper bounds them in terms of goodput.

*1) Infinite Buffer Experiments:* For the first set of experiments, we assumed an infinitely back-logged queue at the transmitter. Unless otherwise noted, the following parameters were used: block size $n = 1$ packet, feedback delay $d = 1$ packet, mean SNR $\mathrm{E}\{\gamma_t\} = 25$ dB, and fading-rate parameter $\alpha = 0.001$. For each channel realization, 200 packets (each consisting of $p = 100$ symbols) were transmitted. The steady-state goodputs reported (per symbol per packet) in the figures were calculated by averaging instantaneous goodputs over the packets in 1000 channel realizations for Figs. 3-4 and 500 channel realizations for Figs. 5-6. To ensure that steady-state performance was reported, the algorithms were initialized at the goodput-maximizing rate for each new channel realization.

Figure 3 plots steady-state goodput as a function of mean SNR $\mathrm{E}\{\gamma_t\}$. To vary $\mathrm{E}\{\gamma_t\}$, we varied the parameter $K$ while keeping $\alpha = 0.01$. The plot shows the greedy algorithm exhibits an increasing gain over the fixed-rate algorithm as mean SNR increases. At low mean SNR, little gain is observed because the optimal constellation size is almost always the smallest one, as can be inferred from Fig. 2. But, at higher mean SNRs, the greedy algorithm performs about 1 dB worse (in SNR) than the causal genie, whereas the fixed-rate scheme performs about 5 dB worse. Furthermore, the SNR gap between the greedy and fixed-rate schemes grows as mean SNR increases. Since the steady goodput achieved by the causal genie upper bounds that achievable by any causal-feedback-based rate adaptation algorithm, one can infer that greedy adaptation based on 1-bit ACK/NAK feedback is sufficient to attain a major fraction of the gain achievable by any causal feedback scheme.

Figure 4 shows steady-state goodput versus fading-rate parameter $\alpha$ for mean SNR $\mathrm{E}\{\gamma_t\} = 25$ dB. Lower $\alpha$ corresponds to slower channel variation and thus more accurate prediction of instantaneous SNR. From the plot, the following can be observed: as $\alpha$ decreases, both the causal genie and the greedy algorithm approach the non-causal genie, whereas as $\alpha$ increases, both the causal genie and the greedy algorithm approach the fixed-rate algorithm. The non-causal genie and fixed-rate algorithms yield essentially constant[9] steady-state goodput versus $\alpha$. For a wide range of $\alpha$, it can be seen that the greedy algorithm performs closer to the causal genie than it does to the fixed-rate algorithm. Thus, we conclude that the greedy scheme captures a dominant fraction (e.g., $\approx 90\%$ at low $\alpha$) of the goodput gain achievable under causal feedback.

Figure 5 plots steady-state goodput versus feedback delay $d$ for packet-rate adaptation, i.e., $n = 1$. By definition, the non-causal genie has access to all past, current, and future SNRs, so its performance is unaffected by delay. As for the causal genie and greedy algorithms, their steady-state goodputs measure 30% and 20% above that of the fixed-rate algorithm, respectively, when $d = 1$. However, as the delay $d$ increases, their causally predicted SNR distributions converge to the prior SNR distribution, so that, the causal genie and greedy algorithms eventually perform no better than the fixed-rate algorithm. Still, for all delays, the simple greedy scheme captures a dominant fraction of the goodput gain achievable under causal feedback.

Figure 6 plots steady-state goodput versus block size $n$ packets for delay $d = 1$ packet and $\alpha = 0.001$. For all tested block sizes, the greedy algorithm performs closer to the causal genie than to the fixed-rate algorithm, implying that the greedy algorithm once again recovers a dominant portion of the goodput gain achievable under the causal feedback constraint. The performances of all adaptive schemes decrease with block size, though. This is for two reasons: first, a uniform rate is applied across the block, whereas the optimal rate varies across the block; and, second, as the block length increases, the SNR must be predicted farther into the future. Notice that even the performance of non-causal genie degrades as $n$ increases due to the sub-optimality of its uniform rate assignment across the block.

Figures 5 and 6 also plot the performance of the so-called *quantized genie* reference scheme, which adapts the rate to maximize goodput under quantized, but otherwise perfect, knowledge of SNR $\gamma_{t-d}$. The goodput attained by the quantized genie upper bounds[10] the goodput attained by a transmitter that assumes a *finite-state* Markov SNR model and employs optimal POMDP-based rate assignment. To construct the corresponding finite-state Markov model, we quantized the SNR using the Lloyd-Max algorithm [23] and calculated the state transition probability matrix via [24, eq. (15)-(16)]. Apart from the finite-state SNR model, rate assignment for the quantized genie is identical to that for the causal genie.

---

[9] Deviations from constant are due to finite averaging effects.

[10] The fact that the quantized genie yields an upper bound in the case of a finite-state Markov channel follows directly from Lemma 1, which holds for both continuous and finite-state Markov channels.

Figures 5 and 6 show that the greedy algorithm outperforms the 2- and 4-state quantized genies, and performs on par with the 7-state quantized genie, throughout most of the examined range of $d$ and $n$. Thus, we conclude that the greedy algorithm outperforms the optimal POMDP-based rate adaptation scheme based on a finite-state Markov SNR model with 7 states or less. This is notable because the computational complexity of optimal POMDP-based rate adaptation is significant, under typical horizons, for channel models with more than a few states.

*2) Finite Buffer Experiments:* For this second set of experiments, a finite data buffer was employed at the transmitter. Bits are removed from the buffer when an ACK arrives, confirming their successful transmission, or when the buffer overflows. The following parameters were used: block size $n = 1$ packet, feedback delay $d = 1$ packet, and mean SNR $\mathrm{E}\{\gamma_t\} = 25$ dB. The packet arrival rate followed a 2-state Markov model with ON and OFF states. In the ON state, a single packet arrives in the buffer (queue), and in the OFF state, no packets arrive. The self transition probability in both ON and OFF states was set to $0.9$ in order to mimic bursty traffic. Consequently, the steady-state probability of each state is $0.5$ and the long-term arrival rate is $0.5$ packets/interval. The size of an arriving packet was set equal to the number of bits transmitted (per packet interval) by the fixed-rate reference scheme under backlogged conditions. The size of the buffer was set equal to 30 such packets of data. Thus, if packets were arriving persistently, then, in the absence of NAKs, the fixed-rate scheme would yield a fixed buffer occupancy, while, in the absence of ACKs, the buffer would go from totally empty to totally full after 30 arrivals. For each channel realization, 1000 packets were transmitted (each consisting of $p = 100$ symbols) and the buffer was initialized at half-full. The values reported in the figures represent the average of all packets in 1000 channel realizations.

Figure 7 plots average buffer occupancy versus fading-rate parameter $\alpha$, where a buffer occupancy of "$b$" is to be interpreted as $b$ arrival-packets worth of bits. It can be seen that the buffer occupancy achieved by the greedy algorithm is very close to that achieved by the causal and non-causal genie algorithms, whereas the buffer occupancy achieved by the fixed-rate scheme is much higher, especially at lower values of $\alpha$. Recall that, when $\alpha$ is low, the SNR can remain below average for prolonged periods of time, during which fixed-rate transmissions are more likely to yield NAKs and hence fill the buffer. Figure 8 plots a related statistic: the fraction of packets that are dropped due to buffer overflows. Here again, the drop rate achieved by the greedy algorithm is very close to that achieved by the causal and non-causal genie algorithms, whereas the drop rate achieved by the fixed-rate algorithm is more than 10 times higher.

Figure 9 shows steady-state goodput versus fading-rate parameter $\alpha$ for Markov arrivals and finite buffer size. The steady-state goodput achieved by the greedy scheme is very close to that of the causal and non-causal genie schemes, whereas the steady-state goodput achieved by the fixed-rate scheme is much lower, especially when $\alpha$ is small. The increase of steady-state goodput with $\alpha$ is directly related to the decrease in drop rate with $\alpha$ observed in Fig. 8, since dropped packets do not contribute to goodput.

## VI. CONCLUSION

In this paper, we studied rate adaptation schemes that use degraded error-rate feedback (e.g., packet-rate ACK/NAKs) to maximize finite-horizon expected goodput over continuous Markov flat-fading wireless channels. First, we specified the POMDP that leads to the optimal rate schedule and showed that its solution is computationally impractical. Then, we proposed a simple greedy alternative and showed that, while generally suboptimal, the greedy approach is optimal when the error-rate feedback is non-degraded. We then detailed an implementation of the greedy rate-adaptation scheme in which the SNR distribution is estimated online (from degraded error-rate feedback) and combined with offline-calculated goodput-versus-SNR curves to find the expected-goodput maximizing transmission rate. In addition to the packet-rate greedy adaptation scheme, a block-rate greedy adaptation scheme was also proposed that offers the potential for significant reduction in complexity with only moderate sacrifice in performance.

For the particular case of uncoded square-QAM transmission, packet-rate ACK/NAK feedback, and Rayleigh fading, the greedy scheme was numerically compared to three reference schemes: the optimal fixed-rate scheme, a genie-aided scheme with perfect causal SNR knowledge, and a genie-aided scheme with perfect non-causal SNR knowledge. First, the effects of mean SNR, channel fading rate, and feedback delay on steady-state goodput were investigated in the context of an infinitely backlogged transmission queue. In this case, the causal genie reference is especially meaningful because it upper bounds the performance of the optimal POMDP scheme, which is too complex to implement directly. Second, a finite transmission buffer was considered, and the effects of channel fading rate on buffer occupancy, drop rate, and steady-state goodput were investigated. The results suggest that the simple packet-rate greedy scheme captures a dominant fraction of the achievable goodput under causal feedback, whereas the optimal fixed-rate scheme captures significantly less. Similarly, the drop rate and average buffer occupancy of the greedy scheme were nearly equal to those of the causal and non-causal genie-aided schemes, whereas the drop rate and average buffer occupancy of the fixed-rate scheme were much higher (e.g., an order-of-magnitude higher in the case of drop rate). Comparisons to a "quantized genie" scheme that upper bounds optimal adaptation under a finite-state Markov SNR model were also made, and there it was found that the proposed greedy scheme outperformed the quantized genie scheme with up to 7 states. Since POMDP-based optimal rate-adaptation for discrete-Markov channels with 7 or more states would be computationally intensive, greedy rate-adaptation based on a continous-Markov channel model is more appealing.

## References

[1] A. Goldsmith, *Wirless Communications*. New York: Cambridge University Press, 2005.

[2] A. Goldsmith and S. Chua, "Variable rate variable power M-QAM for fading channels," *IEEE Trans. Commun.*, vol. 45, pp. 1218–1230, Oct. 1997.

[3] D. L. Goeckel, "Adaptive coding for time-varying channels using outdated fading estimates," *IEEE Trans. Commun.*, vol. 47, pp. 844–855, June 1999.

[4] K. Balachandran, S. R. Kadaba, and S. Nanda, "Channel quality estimation and rate adaptation for cellular mobile radio," *IEEE J. Select. Areas In Commun.*, vol. 17, pp. 1244–1256, July 1999.

[5] G. Holland, N. Vaidya, and P. Bahl, "A rate adaptive MAC protocol for multi-hop wireless networks," in *Proc. ACM Internat. Conf. on Mobile Computing and Networking*, vol. 5, 2001.

[6] B. Sadegi, V. Kanodia, A. Sabharwal, and E. Knightly, "Opportunistic media access for multirate ad hoc networks," in *Proc. ACM Internat. Conf. on Mobile Computing and Networking*, vol. 5, (Atlanta, GA), pp. 3246–3250, 2001.

[7] J. C. Bicket, "Bit-rate selection in wireless networks," Master's thesis, Massachusetts Institute of Technology, Feb 2005.

[8] S. Wong, H. Yang, S. Lu, and V. Bharghavan, "Robust rate adaptation for 802.11 wireless networks," in *Proc. ACM Internat. Conf. on Mobile Computing and Networking*, 2006.

[9] M. Rice and S. B. Wicker, "Adaptive error control for slowly varying channels," *IEEE Trans. Commun.*, vol. 42, pp. 917–925, Feb./Mar./Apr. 1994.

[10] Y.-D. Yao, "An effective go-back-N ARQ scheme for variable-error-rate channels," *IEEE Trans. Commun.*, vol. 43, pp. 20–23, Jan. 1995.

[11] S. S. Chakraborty, M. Liinabarja, and E. Yli-Juuti, "An adaptive ARQ scheme with packet combining for time varying channels," *IEEE Commun. Letters*, vol. 3, pp. 52–54, Feb. 1999.

[12] S. Choi and K. G. Shin, "A class of hybrid ARQ schemes for wireless links," *IEEE Trans. Veh. Tech.*, vol. 50, pp. 777–790, May 2001.

[13] H. Minn, M. Zeng, and V. K. Bhargava, "On ARQ scheme with adaptive error control," *IEEE Trans. Veh. Tech.*, vol. 50, pp. 1426–1436, Nov. 2001.

[14] A. K. Karmokar, D. V. Djonin, and V. K. Bhargava, "POMDP-based coding rate adaptation for Type-I hybrid ARQ systems over fading channels with memory," *IEEE Trans. Wireless Commun.*, vol. 5, pp. 3512–3523, Dec. 2006.

[15] D. V. Djonin, A. K. Karmokar, and V. K. Bhargava, "Joint rate and power adaptation for type-I hybrid ARQ systems over correlated fading channels under different buffer-cost constraints," *IEEE Trans. Veh. Tech.*, vol. 57, pp. 421–435, Jan. 2008.

[16] G. E. Monahan, "A survey of partially observable Markov decision processes: Theory, models, and algorithms," *Management Science*, vol. 28, pp. 1–16, Jan. 1982.

[17] C. C. Tan and N. C. Beaulieu, "On first-order Markov modeling for the Rayleigh fading channel," *IEEE Trans. Commun.*, vol. 48, pp. 2032–2040, Dec. 2000.

[18] J. G. Proakis, *Digital Communications*. New York: McGraw-Hill, 3rd ed., 1995.

[19] D. Bertsekas, *Dynamic Programming and Optimal Control*. Athena Scientific, 2nd ed., 2000.

[20] C. H. Papadimitriou and J. N. Tsitsiklis, "The complexity of Markov decision processes," *Mathematics of Operations Research*, vol. 12, no. 3, pp. 441–450, 1987.

[21] H. V. Poor, *An Introduction to Signal Detection and Estimation*. New York: Springer, 2nd ed., 1994.

[22] I. Abou-Fayal, M. Medard, and U. Madhow, "Binary adaptive coded pilot symbol assisted modulation over rayleigh fading channels without feedback," *IEEE Trans. Commun.*, vol. 53, pp. 1036–1046, June 2008.

[23] A. Gersho and R. M. Gray, *Vector Quantization and Signal Compression*. Boston: Kluwer Academic Publishers, 1992.

[24] P. Sadeghi, R. Kennedy, P. Rapajic, and R. Shams, "Finite-state markov modeling of fading channels - a survey of principles and applications," *IEEE Signal Processing Mag.*, vol. 25, pp. 57–80, Sept. 2008.

## Appendix

Here, we derive the expression for $p(\gamma_t \mid \gamma_{t-nd})$ given in (44). Let $g_{t,R}$ and $g_{t,I}$ be the real and imaginary parts of channel gain, $g_t$. Also let $g_{t-nd} = |g_{t-nd}|e^{j\theta}$ for $\theta \sim \mathrm{U}(0, 2\pi)$. Then

$$p(\gamma_t \mid \gamma_{t-nd}) = \int_0^{2\pi} p\left(\gamma_t \mid \gamma_{t-nd}, \theta\right) p(\theta) d\theta. \tag{45}$$

We first find $p(|g_t| \mid \gamma_{t-nd}, \theta)$ in order to evaluate $p(|g_t| \mid \gamma_{t-nd})$. Since

$$g_t = (1 - \alpha)^{nd}|g_{t-nd}|e^{j\theta} + Z \tag{46}$$

for $Z = \alpha \sum_{i=0}^{nd-1}(1-\alpha)^j w_{t-j}$ and $|g_t| = \sqrt{\frac{\gamma_t}{K}}$, then, conditional on the pair $(\gamma_{t-nd}, \theta)$, the random variables $g_{t,R}$ and $g_{t,I}$ are both Gaussian with mean

$$\mathrm{E}\{g_{t,R} \mid \gamma_{t-nd}, \theta\} = (1 - \alpha)^{nd}\sqrt{\frac{\gamma_{t-nd}}{K}}\cos\theta \tag{47}$$

and

$$\mathrm{E}\{g_{t,I} \mid \gamma_{t-nd}, \theta\} = (1 - \alpha)^{nd}\sqrt{\frac{\gamma_{t-nd}}{K}}\sin\theta, \tag{48}$$

respectively, and variance $\sigma_Z^2 = \mathrm{E}\{Z^2\}$. Thus conditional on $(\gamma_{t-nd}, \theta)$, the random variable $|g_t| = g_{t,R}^2 + g_{t,I}^2$ is Rician [1, p. 78]:

$$
\begin{aligned}
p(|g_t| \mid \gamma_{t-nd}, \theta) = {} & \frac{|g_t|}{\sigma_Z^2} \exp\left(\frac{-\left(|g_t|^2 + (1-\alpha)^{2nd}\frac{\gamma_{t-nd}}{K}\right)}{2\sigma_Z^2}\right) \\
& \times I_0\left(\frac{|g_t|(1-\alpha)^{nd}\sqrt{\frac{\gamma_{t-nd}}{K}}}{\sigma_Z^2}\right).
\end{aligned} \tag{49}
$$

One can see that, given $\gamma_{t-nd}$, the random variable $|g_t|$ is independent of $\theta$. Since $\gamma_t = K|g_t|^2$, we have

$$
\begin{aligned}
p(\gamma_t \mid \gamma_{t-nd}) = {} & \frac{1}{2K\sigma_Z^2} \exp\left(\frac{-\left(\frac{\gamma_t}{K} + (1-\alpha)^{2nd}\frac{\gamma_{t-nd}}{K}\right)}{2\sigma_Z^2}\right) \\
& \times I_0\left(\frac{(1-\alpha)^{nd}\sqrt{\gamma_t\gamma_{t-nd}}}{K\sigma_Z^2}\right).
\end{aligned} \tag{50}
$$

Hence combining (45) and (50), we get

$$
\begin{aligned}
p(\gamma_t \mid \gamma_{t-nd}) = {} & \frac{1}{2K\sigma_Z^2} \exp\left(\frac{-\left(\gamma_t + (1-\alpha)^{2nd}\gamma_{t-nd}\right)}{2K\sigma_Z^2}\right) \\
& \times I_0\left(\frac{(1-\alpha)^{nd}\sqrt{\gamma_t\gamma_{t-nd}}}{K\sigma_Z^2}\right).
\end{aligned} \tag{51}
$$

PSfrag replacements

Finally, plugging $\sigma_Z^2 = \frac{\alpha}{2-\alpha}\left(1 - (1-\alpha)^{2nd}\right)$ into (51) yields (44).

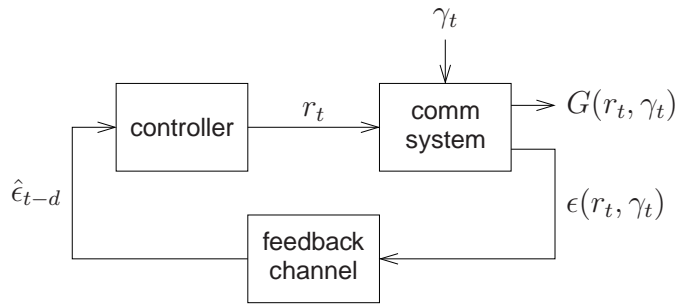

Fig. 1. System model.

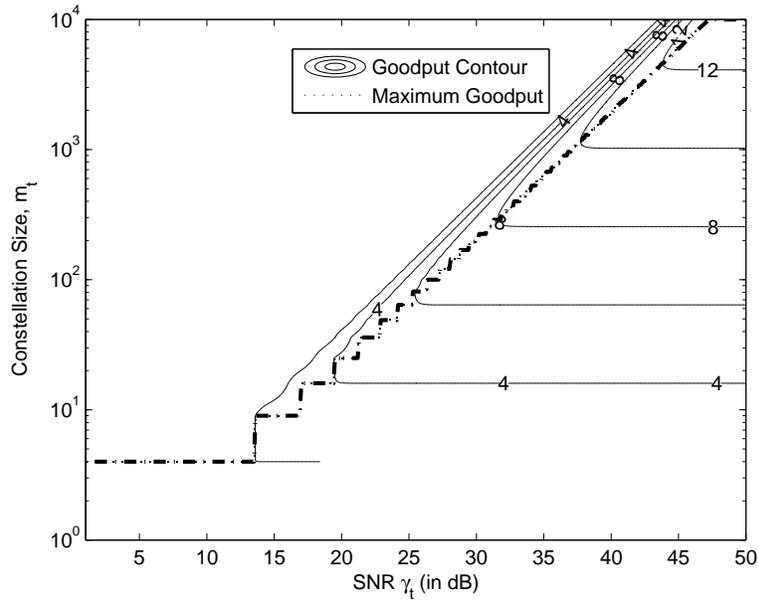

Fig. 2. Goodput contours versus SNR $\gamma_t$ and constellation size $m_t$ for packet size $p = 100$. The goodput maximizing constellation size, as a function of SNR, is shown by the dash-dot line.

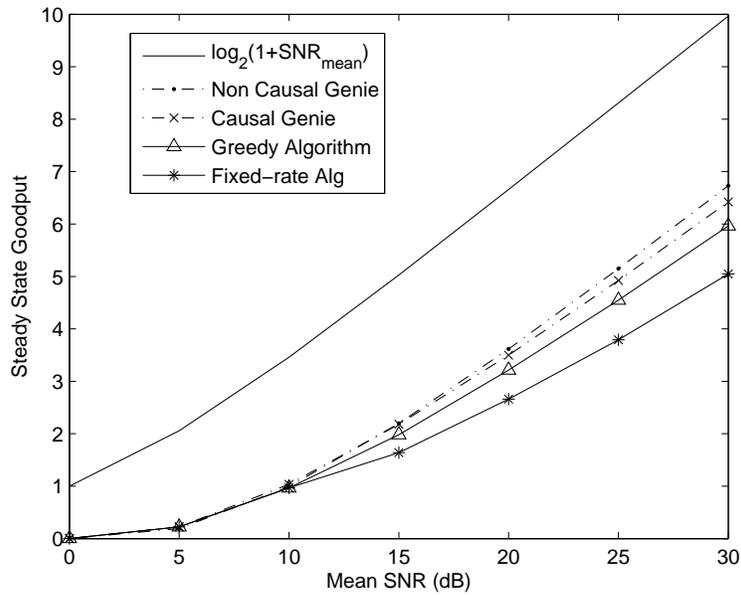

Fig. 3. Steady-state goodput versus mean SNR $\mathrm{E}\{\gamma_t\}$ for $\alpha = 0.01$, block size $n = 1$ packet, and delay $d = 1$ packet.

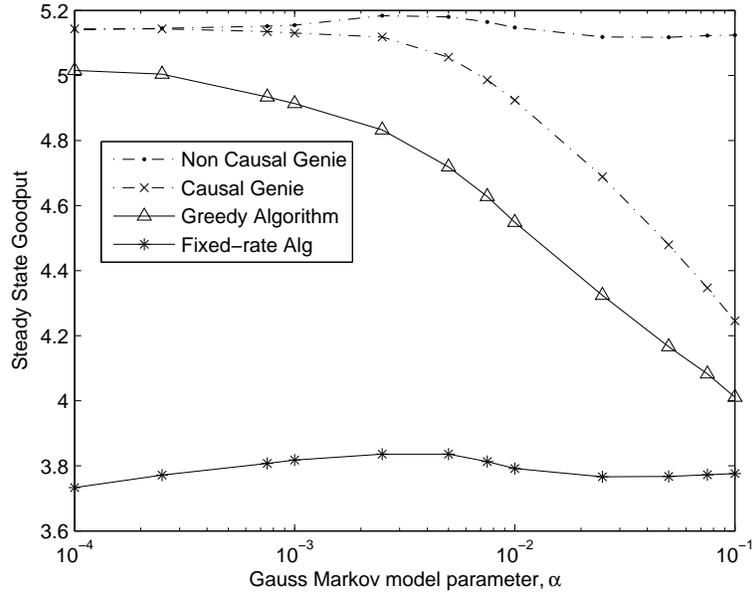

Fig. 4. Steady-state goodput versus $\alpha$ for $\mathrm{E}\{\gamma_t\} = 25$ dB, block size $n = 1$ packet, and delay $d = 1$ packet.

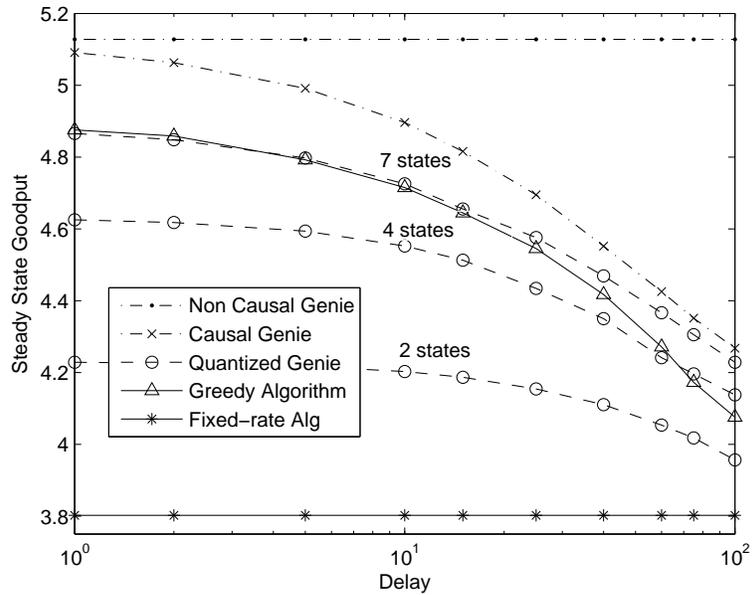

Fig. 5. Steady-state goodput versus delay $d$ for $\mathrm{E}\{\gamma_t\} = 25$ dB, $\alpha = 0.001$, and block size $n = 1$ packet.

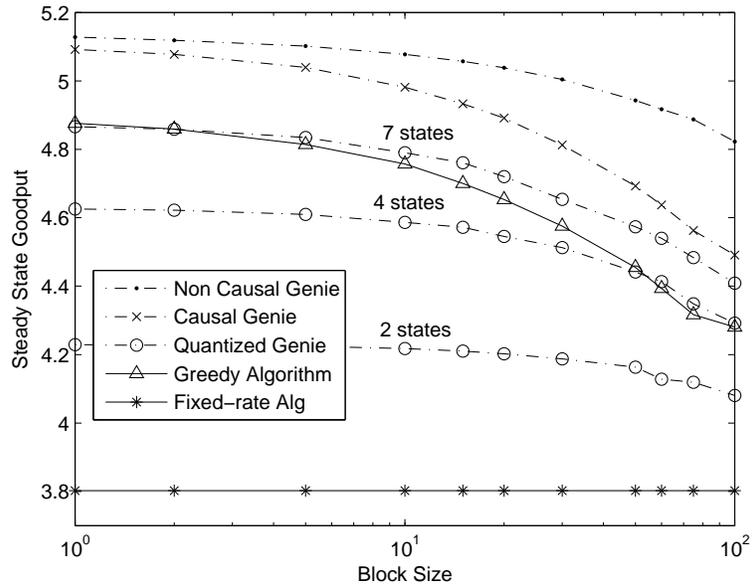

Fig. 6.   Steady-state goodput versus block size $n$ for $\mathrm{E}\{\gamma_t\} = 25$ dB, $\alpha = 0.001$, and delay $d = 1$ packet.

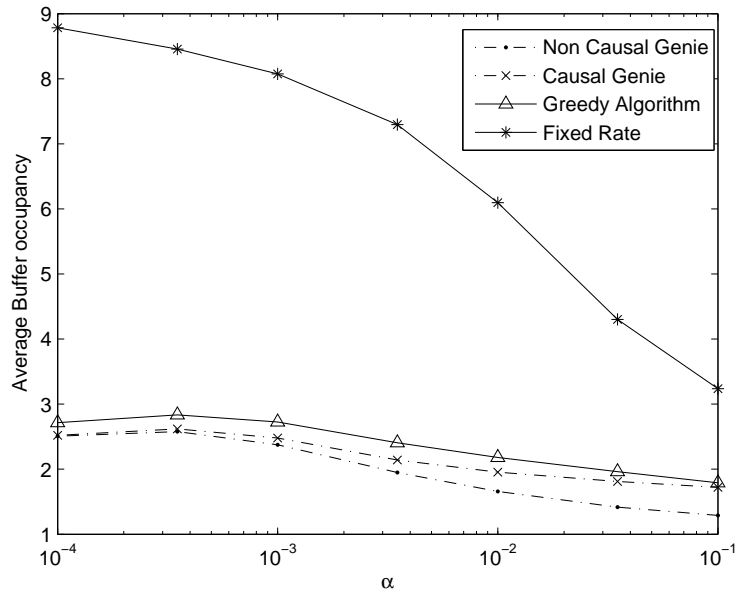

Fig. 7.   Average buffer occupancy versus $\alpha$ for Markov arrivals with average rate = 0.5 packets/interval, buffer size = 30 packets, $\mathrm{E}\{\gamma_t\} = 25$ dB, block size $n = 1$ packet, and delay $d = 1$ packet.

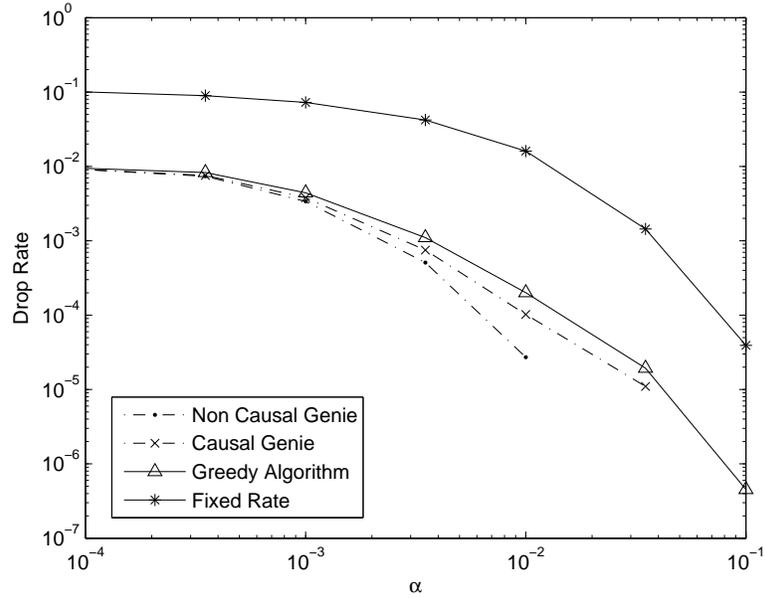

Fig. 8.  Average drop rate versus $\alpha$ for Markov arrivals with average rate = 0.5 packets/interval, buffer size = 30 packets, $\mathrm{E}\{\gamma_t\} = 25$ dB, block size $n = 1$ packet, and delay $d = 1$ packet.

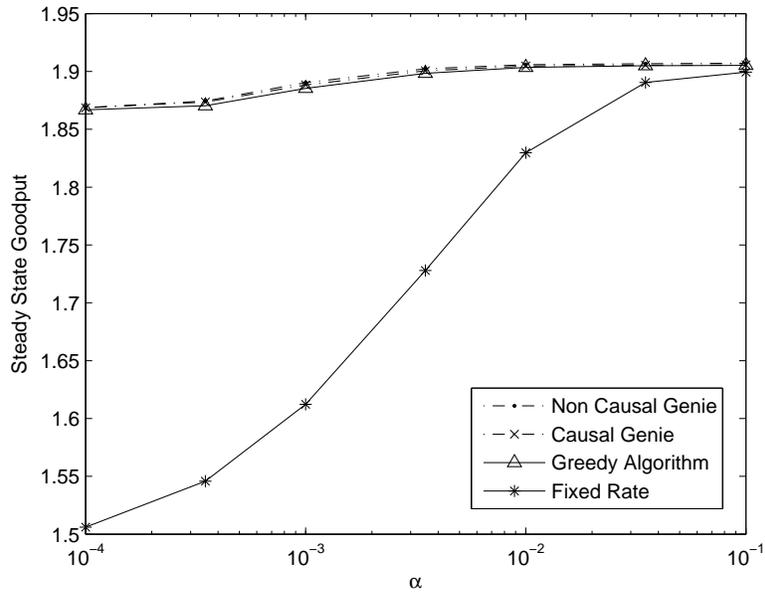

Fig. 9.  Steady-state goodput rate versus $\alpha$ for Markov arrivals with average rate = 0.5 packets/interval, buffer size = 30 packets, $\mathrm{E}\{\gamma_t\} = 25$ dB, block size $n = 1$ packet, and delay $d = 1$ packet.